# A multilevel nonvolatile magnetoelectric memory based on memtranstor

Jianxin Shen, Junzhuang Cong, Dashan Shang, Yisheng Chai, Shipeng Shen, Kun Zhai, and Young Sun

Beijing National Laboratory for Condensed Matter Physics, Institute of Physics, Chinese Academy of Sciences, Beijing 100190, P. R. China

Corresponding author: Y.S. (email: youngsun@iphy.ac.cn)

The coexistence and coupling between magnetization and electric polarization in multiferroic materials provide extra degrees of freedom for creating next-generation memory devices. A variety of concepts of multiferroic or magnetoelectric memories have been proposed and explored in the past decade. Here we propose a new principle to realize a multilevel nonvolatile memory based on the multiple states of the magnetoelectric coefficient ($\alpha$) of multiferroics. Because the states of $\alpha$ depends on the relative orientation between magnetization and polarization, one can reach different levels of $\alpha$ by controlling the ratio of up and down ferroelectric domains with external electric fields. Our experiments in a device made of the PMN-PT/Terfenol-D multiferroic heterostructure confirm that the states of $\alpha$ can be well controlled between positive and negative by applying selective electric fields. Consequently, two-level, four-level, and eight-level nonvolatile memory devices are demonstrated at room temperature. This kind of multilevel magnetoelectric memory retains all the advantages of ferroelectric random access memory but overcomes the drawback of destructive reading of polarization. In contrast, the reading of $\alpha$ is nondestructive and highly efficient in a parallel way, with an independent reading coil shared by all the memory cells.

Ferroelectric random access memory (FeRAM) that stores information by using the spontaneous polarization ($P$) of ferroelectrics is a mature and promising non-volatile memory because of its high endurance, fast read/write speed, low power consumption, and reliable multilevel polarization states[1-5]. Nonetheless, one major problem associated with conventional FeRAM is on the reading operation. The reading of $P$ is usually performed by applying a bias voltage to the ferroelectric capacitor and detecting the $P$ switching current. This process is destructive and a rewrite step is necessary. Moreover, this reading method also requires a minimum capacitor size to generate enough current for the sensing circuit, and thus the storage density is limited. Recently, the photovoltaic effect of ferroelectrics such as $BiFeO_3$ has been used to read the $P$ direction nondestructively[6]. However, this method is restricted to specific ferroelectric materials with a band gap in the visible range and requires additional light source. Here, we propose and demonstrate that the problem of destructive reading of $P$ can be overcome by employing the magnetoelectric (ME) effects of multiferroic materials, which yields a new type of multilevel nonvolatile magnetoelectric memory.

In previously known nonvolatile memories such as magnetic random access memory (MRAM)[7-10], resistive switching random access memory (RRAM)[11,12], phase change memory (PRAM)[13,14], FeRAM[1-6], and multiferroic or magnetoelectric random access memory (MERAM)[15-23], the digital information is usually stored by three quantities, respectively: the direction of $M$ or $P$ and the level of resistance ($R$). In addition to $M$ and $P$, multiferroic materials that combine magnetism and ferroelectricity[24-28], have a unique physical quantity − the ME coefficient ($\alpha$), defined by $\alpha_D = dP/dH$ and $\alpha_C = \mu_0\,dM/dE$, where $H$ is magnetic field, and $E$ is electric field. The former is called the direct ME effect and the latter is called the converse ME effect. Both $\alpha_D$ and $\alpha_C$ can be either positive or negative, depending on the ME coupling mechanism and the status of $M$ and $P$. Therefore, instead of using $M$ and $P$ themselves to store information, one can employ the ME coefficient to encode digital data. As we demonstrate below, this new concept of memory has many advantages over previously proposed multiferroic memories as well as FeRAM.

## Results

**Principle and structure of the memory device.** For the benefit of easy operations in writing and reading, we consider a multiferroic material (either single-phase or composite) with in-plane $M$ and vertical $P$. This multiferroic material is sandwiched between two electrodes to form a memory cell, as shown in Fig. 1. Due to the ME effects in the multiferroic material, a change of in-plane $M$ would induce a change in vertical $P$, and vice versa. When the direction of $M$ remains unchanged, the sign of $\alpha_D = dP/dH$ depends on the direction of $P$: $\alpha_D > 0$ for $+P$ (up) and $\alpha_D < 0$ for $-P$ (down). Therefore, we can set $\alpha_D > 0$ as digital "0" and $\alpha_D < 0$ as "1". The digital information is written by applying an electric field between two electrodes to switch the direction of $P$, like that in FeRAM. To read out information, one simply measures $\alpha_D$ by applying a low $H$ and detecting the induced change of $P$. In practice, the ME voltage coefficient $\alpha_E = dE/dH \propto \alpha_D$, is measured by applying a low magnetic field ($\Delta H$) and detecting the induced voltage ($\Delta V$) between two electrodes (Fig. 1e) − a technique that has been widely used in the study of multiferroic materials[24-26].

In addition to the conventional two states of polarization (up and down), ferroelectrics have an internal degree of freedom that allows multilevel polarization (MLP) states[5]. As shown in Figure 1b and 1d, one can obtain intermediate $P$ states between two saturation values ($+P_S$ and $−P_S$) by adjusting the ratio of up and down ferroelectric domains. Corresponding to different levels of $P$, the ME coefficient $\alpha_E$ also has different values when the direction of $M$ remains unchanged. Thus, by employing the ME coefficient to store digital data, we can realize not only two-state but also multilevel non-volatile memories.

To testify the feasibility of above principle, we have performed experiments in a device made of PMN-PT(110)/Terfenol-D heterostructure. PMN-PT is a well-known ferroelectric with a large piezoelectric effect and Terfenol-D is a famous magnetostrictive material. Thus, this typical multiferroic heterostructure has pronounced ME coupling via the interfacial strain. Figure 2a plots the structure of the device and the configuration of measurements. The $E$ field is applied between two Ag electrodes to switch the direction of $P$ in PMN-PT layer and the dc magnetic field $H_{dc}$ is applied in plane to control the $M$ of Terfenol-D layer. For the measurement of $\alpha_E$, a conventional dynamic technique is used, where a small ac magnetic field $h_{ac}$ is applied in plane to induce a voltage change between electrodes via the ME effects.

**Performance of the memory device.** Figure 2b shows how the ME voltage coefficient $\alpha_E$ of the device depends on the status of both $M$ and $P$. Before measuring $\alpha_E$, the device was pre-poled by applying a positive or a negative $E$ field of 4 kVcm$^{-1}$ to set the direction of $P$. Then, $\alpha_E$ was measured as a function of in-plane $H_{dc}$. When $P$ is set to up (the red curve), $\alpha_E$ is very small in the high $H_{dc}$ region because $M$ is saturated and the magnetostriction coefficient is nearly zero. As $H_{dc}$ decreases from 10 kOe to zero, $\alpha_E$ increases steadily and exhibits a maximum around 1 kOe where the magnetostriction coefficient of Terfenol-D is largest. When $H_{dc}$ scans from positive to negative, $\alpha_E$ also changes its sign from positive to negative and shows a minimum around −1 kOe. In contrast, when $P$ is set to down (the black curve), the $H$ dependence of $\alpha_E$ is totally opposite, being negative for $+H_{dc}$ and positive for $−H_{dc}$. These results reveals that the sign of $\alpha_E$ depends on the relative orientation between $M$ and $P$: when the direction of $P$ is fixed, the sign of $\alpha_E$ can be switched by reversing $M$ with $H_{dc}$; when the direction of $M$ is fixed, the sign of $\alpha_E$ can be switched by reversing $P$ with external $E$. The latter case is employed for the memory device because it is more convenient to apply $E$ field in electric circuit.

Though $\alpha_E$ has a maximum with a dc bias field $H_{dc}$ around 1 kOe, it is unfavorable to apply a bias field in practical devices. Thanks to the hysteresis in Fig. 2(b), there is a remanence of $\alpha_E$ at zero bias field. In the following, we focus on the performance of the memory device without a dc bias field.

Firstly we demonstrate the two-state non-volatile memory using $\alpha_E$ of the device. After applying an $E$ pulse of +4 kVcm$^{-1}$ to set polarization to +$P_s$, $\alpha_E$ was measured for 100 s; then, another $E$ pulse of −4 kVcm$^{-1}$ was applied to reverse +$P_s$ to − $P_s$ and $\alpha_E$ was measured for 100 s again. This process was repeated for many cycles. As seen in Fig. 3a, once the applied $E$ pulse reverses $P$, $\alpha_E$ inverts its sign and retains the state until next $E$ pulse is applied. Consequently, $\alpha_E$ switches between positive and negative periodically with external $E$ pulses. When we reduce the negative $E$ field from −4 kVcm$^{-1}$ to −3 kVcm$^{-1}$, only a small portion of ferroelectric domains are revered. As a result, $\alpha_E$ does not changes its sign to negative but retains a positive low value (Fig. 3b). In this case, $\alpha_E$ switches repeatedly between high and low levels, like the resistance switch in RRAM and PCRAM.

Besides the conventional two-state memory, we can achieve multilevel ($2^n$) non-volatile switch of $\alpha_E$ by carefully selecting the amplitude of -$E$ field. Figure 4a and 4b demonstrate the four-level ($2^2$) and eight-level ($2^3$) switch of $\alpha_E$, respectively. Initially, we apply a +4 kVcm$^{-1}$ $E$ pulse to set $\alpha_E$ to the positive maximum. Starting from this initial state, we can reach different levels of $\alpha_E$ by applying selective –$E$ pulse to either fully or partially reverse the ferroelectric domains. After each $E$ pulse, $\alpha_E$ remains its state without apparent decay. Therefore, these well separated levels of $\alpha_E$ ranging from positive to negative constitute an ideal multilevel non-volatile memory. We note that a reset step by applying +4 kVcm$^{-1}$ $E$ pulse is required before writing to different levels. This is to refresh the initial state to ensure that the finial state reached by every selective –$E$ pulse is reproducible.

**Discussion**

The above experiments confirm the feasibility of our new principle for multilevel non-volatile memory: the states of the ME coefficient $\alpha$ can be effectively employed to store digital information. Comparing with other known non-volatile memories, this kind of magnetoelectric memory based on the ME coefficient has many advantages. Firstly, the memory cell has a simple structure so that it is easy to fabricate. Secondly, the writing operation is electrically and fast, by applying a voltage pulse between two electrodes, just like that in FeRAM and RRAM. Thirdly, the reading operation is much easier than that in conventional FeRAM because it avoids the destructive reading of $P$. Instead, the information is read out by simply measuring the induced voltage across the electrodes while an independent coil supplies a small $H$. Since the reading of ME voltage does not require a minimum area of memory cell, the storage density is not restricted as in conventional FeRAM. Moreover, as illustrated in Fig. 1e, all the memory cells can share a single reading coil and all the stored information can be read out in a parallel way. This would greatly simplify the fabrication and operation of the whole memory device. Fourthly, as the device is made of insulating multiferroics and both the writing and reading operations avoid considerable currents, it has a very low power consumption. Finally, compared with the multiple states of resistance in conventional multilevel memories, the multiple levels of the ME coefficient are more distinguishable because it ranges from positive to negative rather than high to low only.

Lastly, we want to point out that the memory device demonstrated in this work has a more fundamental meaning in science: it is regarded as the fourth memelement, in addition to memristor, memcapacitor, and meminductor. As we discussed in ref. 29 and 30, the memory cell exhibiting a pinched or butterfly-shaped hysteresis loop of $\varphi-q$ relationship via the nonlinear ME effects is termed memtranstor. Just like the memristor that has potential applications in memory devices, the memtranstor also has a great promise in developing next-generation memory devices.

The principle of employing the ME coefficient of memtranstors to store digital information paves a new pathway towards an ultimate memory. Besides the PMN-PT/Terfenol-D heterostructure used in this work, a variety of multiferroic heterostructures and even single-phase multiferroic materials could be used as the memory cell once their ME coefficients can be reliably measured. In the future, a systematical investigation on the selection of materials and structures to optimize the performance of memtranstors for industry applications is highly required.

## Methods

**Fabrication of the memory device.** The ferroelectric/ferromagnet multiferroic device was prepared by using $0.7Pb(Mg_{1/3}Nb_{2/3})O_3$-$0.3PbTiO_3$ (PMN-PT) single-crystal substrate with (110)-cut and $Tb_{0.28}Dy_{0.72}Fe_{1.95}$ (Terfenol-D) polycrystalline alloy. Two thin plates of Terfenol-D and PMN-PT ($2 \text{ mm} \times 2 \text{ mm} \times 0.2 \text{ mm}$) were bonded together by using a silver epoxy (Epo-Tek H20E, Epoxy Technology Inc.) to form a multiferroic heterostructure. The top and bottom sides of the structure were covered with sliver paint (Structure Probe, Inc.) to act as the electrodes.

**Measurement of the ME voltage coefficient $\alpha_E$.** A conventional dynamic technique was employed to measure the ME coefficient. A Keithley 6221 AC source was used to supply an ac current to the solenoid to generate a small ac magnetic field $h_{ac}$ at a frequency of 100 kHz. In response, a synchronized 100 kHz ac ME voltage, $V_{ac}=x+y$i, across the electrodes was measured by a lock-in amplifier (Stanford Research SR830). The ME voltage coefficient $\alpha_E$ is calculated by $\alpha_E = x/(h_{ac}t)$, where $t$ is the thickness of the ferroelectric layer. To switch the electric polarization of PMN-PT, a Keithley 6517B electrometer was used to apply voltage pulse across the electrodes. The device was loaded in an Oxford TeslatronPT superconducting magnet system to apply the dc bias magnetic field ($H_{dc}$). All the measurements were performed at 300 K.

**Acknowledgments**

This work was supported by the National Natural Science Foundation of China (Grant Nos. 11227405, 11534015, 51371192, 11274363), and the Chinese Academy of Sciences (Grants No. XDB07030200 and KJZD-EW-M05).

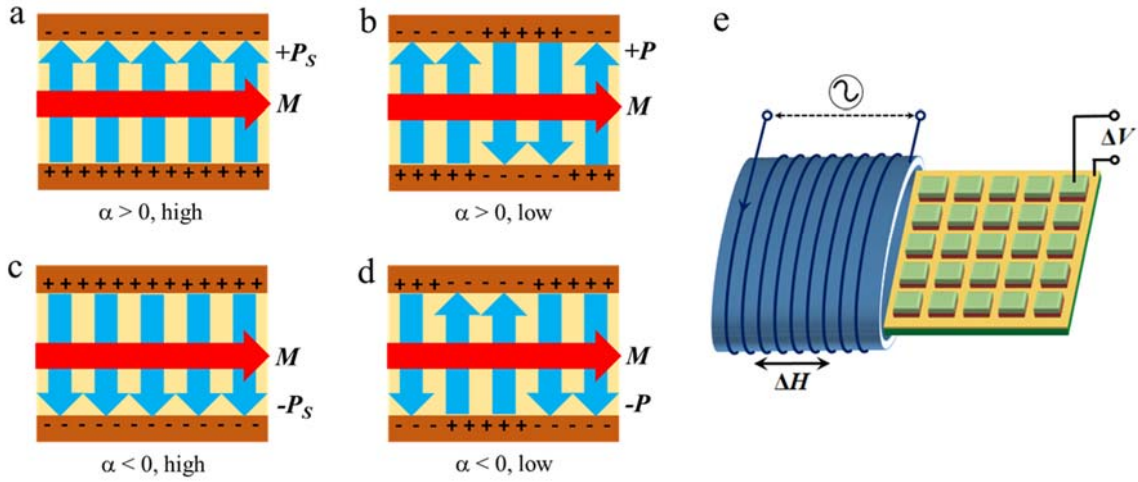

**Figure 1. The principle of the multilevel nonvolatile memory.** (a)-(d) The schematic of a memory cell showing different states of the ME coefficient $\alpha$. When the direction of $M$ remains unchanged, the value of $\alpha$ depends on the ratio of up and down ferroelectric domains, ranging from positive to negative. (e) The illustration of reading operation. An independent reading coil generates a small ac magnetic field $\Delta H$ and the stored information is read out by measuring the induced voltage $\Delta V$. All the memory cells can share a single reading coil, which greatly simplifies the fabrication and operations of the memory device.

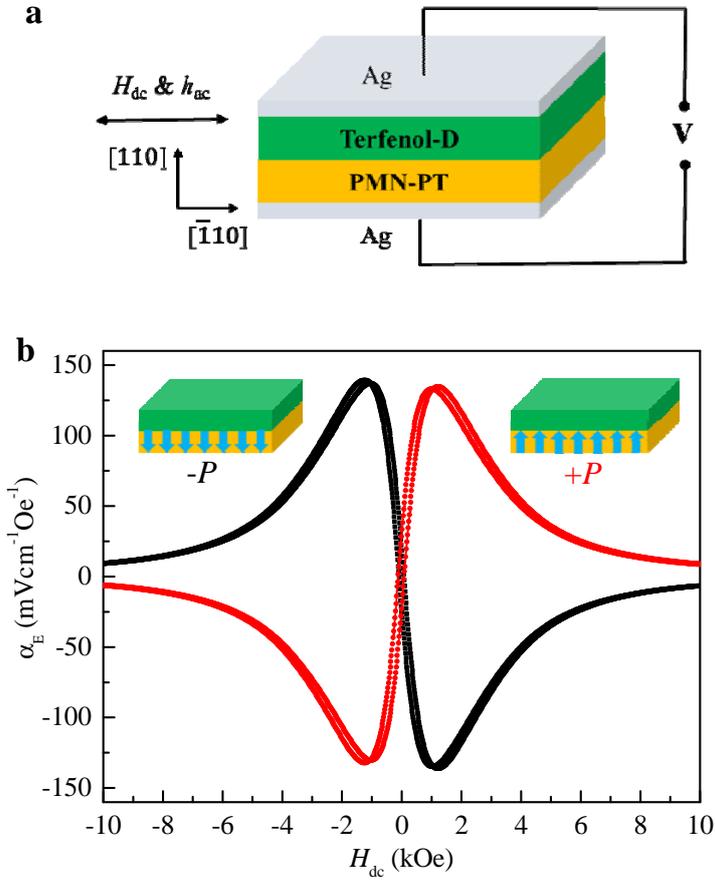

**Figure 2. The ME voltage coefficient $\alpha_E$ of a memory device made of the PMN-PT/Terfenol-D multiferroic heterostructure.** (a) The structure of the device and the measurement configuration. The electric field is applied vertically along [110] of PMN-PT and both the dc bias and ac magnetic fields are applied in plane along [-110] of PMN-PT. (b) $\alpha_E$ as a function of dc bias magnetic field with $+P_s$ and $-P_s$, respectively. The state of $\alpha_E$ depends on the relative orientation between $M$ and $P$, with the maximums located around $\pm 1$ kOe where the magnetostriction coefficient of Terfenol-D is largest.

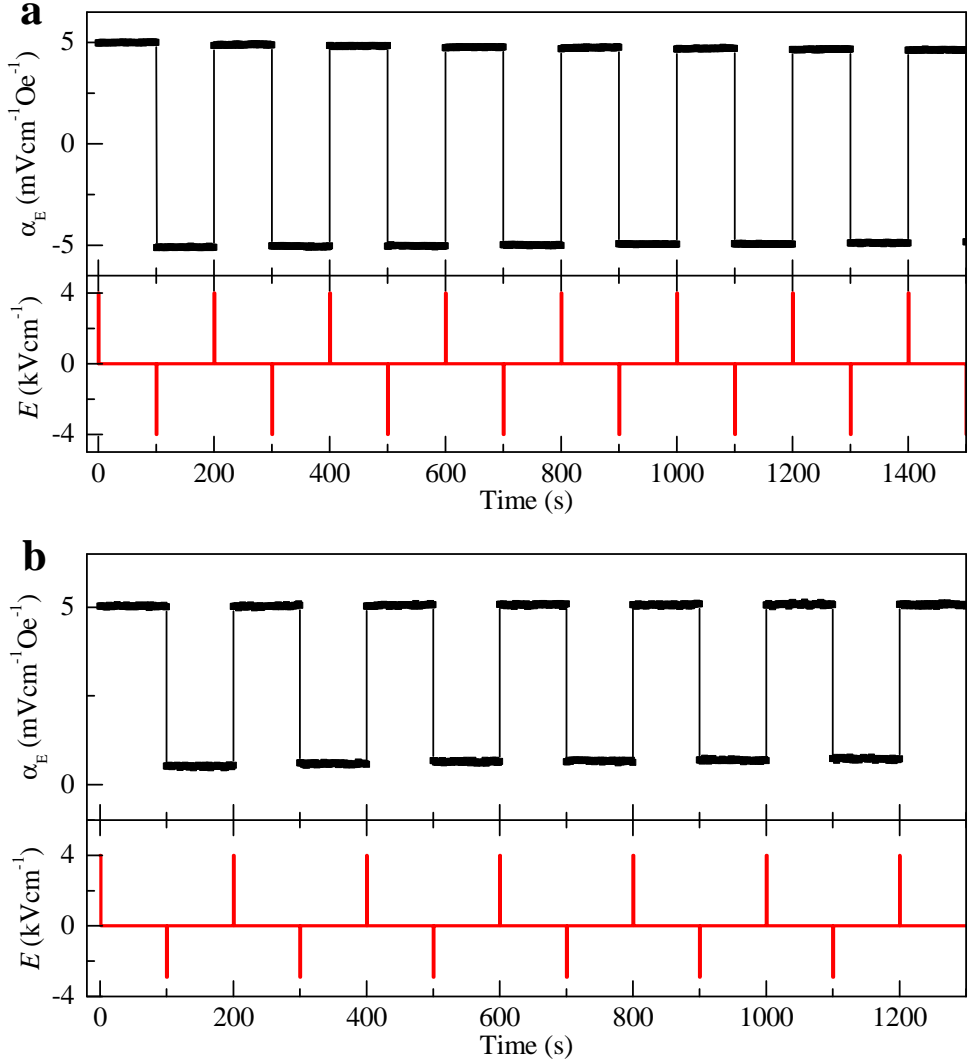

**Figure 3. Demonstration of two-level memory.** (a) Two-level switch of $\alpha_E$ between positive and negative. $E$ pulses of ±4 kVcm$^{-1}$ were applied to fully reverse polarization between +$P_s$ and −$P_s$ so that $\alpha_E$ switches between positive and negative. (b) Two-level switch of $\alpha_E$ between high and low values. $E$ pulse of +4 kVcm$^{-1}$ was used to reach +$P_s$ and -3 kVcm$^{-1}$ was applied to partially reverse ferroelectric domains. As a result, $\alpha_E$ switches between high and low values.

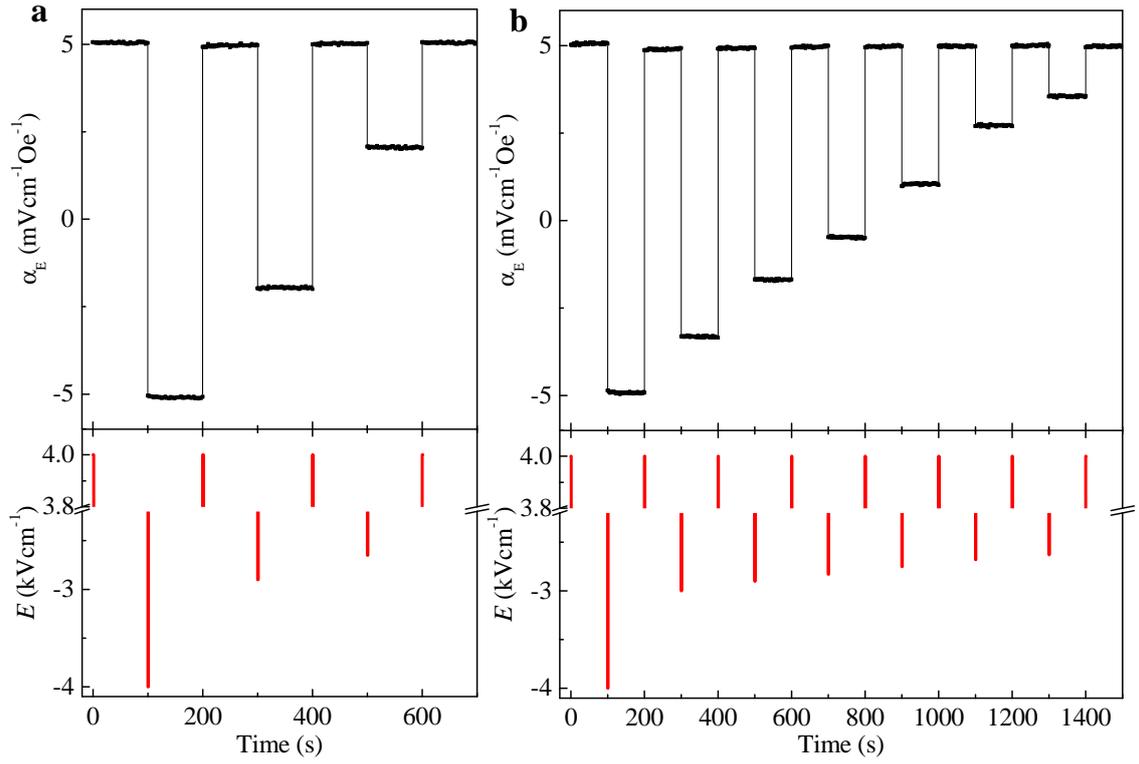

**Figure 4. Demonstration of multilevel memory.** (a) Four-level switch of $\alpha_E$. (b) Eight-level switch of $\alpha_E$. A reset step with +4 kVcm⁻¹ $E$ pulse is performed to refresh the initial state to $+P_s$ before a selective -$E$ pulse is applied to reach different levels of $\alpha_E$.